# Universal Non-Landau, Self-Organized, Lattice Disordering Percolative Dopant Network Sub-T$_c$ Phase Transitions in Ceramic Superconductors

J. C. Phillips

Dept. of Physics and Astronomy, Rutgers University, Piscataway, N. J., 08854

**Abstract**

Ceramic superconductors (cuprates, pnictides, …) exhibit universal features in both $T_c^{max}$ and in their planar lattice disordering measured by EXAFS, as reflected by three phase transitions. The two highest temperature transitions are known to be associated with formation of pseudogaps and superconductive gaps, with corresponding Landau order parameters, but no new gap is associated with the third transition below $T_c$. It is argued that the third transition is a dopant glass transition, which is remarkably similar to transitions previously observed in chalcogenide and oxide alloy network glasses (like window glass).

**1. Introduction**

Conventional metals achieve high superconductive transition temperatures by means of strong electron-phonon interactions which provide an attractive interaction for Cooper pairs that overcomes Coulomb repulsion. In ceramic materials (oxides, halides, pnictides) one would expect to find strong ionic charges, little screening, and correspondingly strong Coulomb interactions, which would make superconductivity unlikely. Moreover, enhancing the electron-phonon interactions usually leads to large Jahn-Teller lattice distortions, which convert metals into semiconductors or insulators. Yet in certain ceramic materials nature has found a way to circumvent both of these unfavorable mechanisms, and to produce not only superconductivity, but even superconductivity at temperatures far higher than are usually found in metals, thus posing the greatest scientific paradox discovered in recent decades [1].

Several theoretical mechanisms propose to resolve this paradox by invoking interactions other than the traditional attractive electron-phonon interactions, but so far there has been little evidence to support such models, apart from the fact that magnetic nanophases often coexist with superconductive nanophases in samples with Meissner filling factors that are never close to unity (and seldom reported). The simplest idea, which recommends itself as minimally complex, is that certain special ceramics (like the layered cuprates) have very strong electron-phonon interactions, and that upon being doped, the resulting structures lose most of their magnetic nanophases, and instead phase-separate into pseudogapped (charge density wave, CDW, or Jahn_Teller distorted) phases and metallic superconductive nanophases [2]. These phases have long been described as "stripes" (dynamical or otherwise), but it is noteworthy that in the rare materials (such as $La_{2-x}Sr_xCuO_4$ near $x = 1/8$) where stripes have been observed by diffraction, $T_c$ has been either depressed, or reduced to 0. It appears that ordering of nanophases into superlattices describable by Landau order parameters is extremely unfavorable for superconductivity.

The opposite of "stripes" (superlattice ordering) would be the very strong disordering found in network glasses, such as window glass. Doped ceramics still have a host lattice structure, but if the dopants themselves do not order, then they can be regarded as forming a glassy network embedded in the host crystal. The average properties of substitutional dopants can be studied by EXAFS, and the average strain fields induced by them in planar Cu-O bonds studied in detail as a function of temperature T. These strain fields exhibit breaks in slope both at the pseudogap $T^*$ and superconductive $T_c$ transition temperatures in ceramic superconductors based on both $CuO_2$ planes [3], and FeAs planes [4]. Since the magnetic properties of Cu and Fe are quite different, this suggests that the minimal model based on strong electron-phonon interactions alone [5] may be a good starting point for building a universal theoretical model for high temperature superconductivity (HTSC) in layered ceramics. This model has recently provided the only known strong least upper bounds on $T_c$ in all ceramic superconductors [6]. Its physical motivation is described in detail in [6] (especially the role played by self-organization). It provides detailed models for many other effects as well (isotope effect, precursive diamagnetic effects at high T, normal-state transport anomalies at even higher T's, etc. [6]). The EXAFS data will enable us to separate the pseudogap (or large Jahn-Teller lattice distortion) and superconductive effects, a separation that has proved difficult both in angle-resolved spectroscopy (ARPES) [7] and large scale scanning tunneling microscope (STM) experiments [8].

## 2. Experiments

The EXAFs experiments [3,4] were done on optimally doped (highest $T_c$) $La_{2-x}Sr_xCuO_4$ (x = 0.15), where the pseudogap, metallic and superconductive effects are closely balanced [6]. The samples were single crystals, and the EXAFS analysis focused on the most accurately measurable length, the planar Cu-O bond length. Also studied were the effects of replacing the Cu with a few % of Ni, Co or Mn. The width of the distribution of planar Cu-O bond lengths was measured by the mean-square relative displacement $\sigma^2$(Cu-O), which may reflect local orthorhombic (double-well) splittings.

The most striking features of the experimental data (Fig. 1) were the breaks in slope of $\sigma^2(T)$ at two temperatures, corresponding to the opening of a pseudogap (at $T = T^*$), and the opening of the superconductive gap (at $T = T_c$). These features were common to all samples, and varied only in magnitude. For T above $T^*$, $\sigma^2$ decreased with decreasing T, but between $T^*$ and $T_c$, $\sigma^2$ increased with decreasing T. Below $T_c$ $\sigma^2$ at first decreased, but near $T_{OP} = 0.8T_c$, $\sigma^2$ unexpectedly reversed and began to increase. The origin of this sharp reversal, which is not associated with a conventional transition temperature or an easily identified order parameter, is mysterious.

As is well known, substitutions of Cu by Ni or Co decrease $T_c$, but surprisingly Mn substitution seems to have little effect on $T_c$. Substitution of Ni or Co at 3% reduces the amplitude of the oscillations above and below $T_c$, while 5% virtually erases them completely. The situation for Mn is quite different, as Mn gradually reduces the amplitiude of the oscillations, which are still present at 5% concentrations. Qualitatively Mn is expected to give rise to different behavior, as it is ~10% larger than Co, Ni or Cu (from MnO and NiO lattice constants, for example). Some of these differences can also arise from crystal field splittings of the d electrons on Co and Ni, while $Mn^{2+}$ is expected to have a half-filled d shell with no associated Jahn-Teller distortions. A more detailed discussion of these chemical effects is given in [3,4].

## 3. Theory

The most important features of the data are the changes in slope of $\sigma^2$ that occur around $T^*$, $T_c$ and $T_{OP} \sim (0.6-0.8)T_c$. The increase as T decreases below $T^*$ is easily understood, as the area associated with the formation of pseudogapped nanodomains with local orthorhombic symmetry will grow with decreasing T, as previously discussed [3,4]; these nanodomains are apparent in STM studies [8]. Below $T_c$, as T decreases $\sigma^2$ drops sharply (see Fig. 2) to a level corresponding to the extrapolation of the data above $T^*$, in other words, the superconductive filaments almost completely disrupt the pseudogapped nanodomains and erase the disorder induced by pseudogap charge density waves. The filaments are still separated by pseudogap walls, but these walls are thin and in them $\sigma^2$ is small: one can suppose that the strain and $\sigma^2$ increase.

The 1989 zigzag filamentary model [9] enables us to understand the reversal of $d\sigma^2/dT$ below $T_{OP}$. As T decreases below $T_c$, the density of filaments increases until at $T = T_{OP}$ the filaments have filled the available volume and begin to overlap. Current flow in these filaments has many hydrodynamic properties, which are useful in explaining diamagnetic "precursive" effects that occur between $T_c$ and $T^*$ [6]. At the junction of two hydrodynamic channels flow is always turbulent, and current fluctuations associated with this turbulence can explain the increase in $\sigma^2$ below $T_{OP}$. See Fig. 3 for examples and discussion.

## 4. Prediction

The explanation for the reversal of $\sigma^2$ below $T_p$ given above describes a new kind of non-Landau topological phase transition associated with filamentary space filling. Because filamentary junctions depend on pair spacing, one can expect that near the metal-insulator transition (which occurs near $x = x_0 = 0.06$ in $La_{2-x}Sr_xCuO_4$), one should find $T_c \sim x - x_0$ and $T_p \sim (x - x_0)^2$, or $T_p/T_c \sim x - x_0$. In this relation could hold between $x = 0.07$ and $x = 0.10$.

## 5. FeAs Ceramic Superconductors

There are many similarities between ceramic superconductors based on both $CuO_2$ planes and FeAs planes, and percolative theory is equally successful in predicting $T_c(<R>)$ as a function of the all-atom, equally weighted average number $<R>$ of Pauling resonating chemical bonds in cuprates [6] and pnictides [10] . One would therefore expect to find many similarities in the disorder $\sigma^2$ in the two (chemically quite different) families. As shown in Fig. 1, all the main features of $\sigma^2$ found in the cuprates (not only T* and $T_c$, but also even $T_{OP}$!) are also found in FeAs disorder [4]. According to the glassy network model [11], the main difference between the two families is that the FeAs materials have $T_c^{max}$ centered near the top of the covalent range, $<R> = 2.5$, while the cuprate center is much lower, at $<R> = 2.0$ [6,10]. This shift in $<R>$ may be related to a shift in $T_{OP}/T_c$ from 0.8 (cuprates) to 0.65 (FeAs), as shown in Fig. 1.

## 6. A Third Phase Transition?

There is little doubt that T* and $T_c$ mark phase transitions, but does $T_{OP}$ also mark a phase transition, and if so, what kind of transition could occur below $T_c$? According to the network model, $T_{OP}$ is the temperature where filaments begin to intersect or cross-link, which will rigidify the filamentary array. Thus we may think of this transition as a kind of glass transition, and look to glass phenomenology for an estimate of $T_{OP}/T_c$ at optimal doping. Indeed glass scientists do estimate that $T_g/T_m$ is about 2/3 in many cases, where $T_g$ is the glass transition temperature, and $T_m$ is the crystalline melting point [12,13].

If one wants a specific example of such glasses, one could consider the system of chalcogenide alloys (As,Ge)Se, where Se chains can be cross-linked by As or Ge [11] or the metallic glasses formed by (Zr,Pt) alloys [14]. Probably Urbach tails provide the best example of filamentary internal ordering of electronic states near a band gap [15]. In all theoretical studies of glasses, the glass transition occurs with decreasing T as the result of increasing stiffness of a space - filling deeply supercooled liquid. The role of network stiffening is particularly clear in the case of the network glasses, where an intermediate phase has been identified, which has many similarities to ceramic superconductive phases [11]. Such comparisons might have seemed remote in 2005, except that now we have identified similar universal features of ceramic superconductors in terms of both a universal master curve for $T_c^{max}$, and here three characteristic transitions, $T^*$, $T_c$ and $T_{OP}$, where only one is seen in traditional metallic superconductors.

I am most grateful to Dr. H. Oyanagi for stimulating correspondence.

## Figure Captions

Fig. 1. This is the inset of Fig. 3 of [4], annotated to emphasize the third phase transition at T = $T_{OP}$ .

Fig. 2. The extrapolation of the high-temperature data (smooth curve) through the $T_c$ peak shows that the superconductive ordering below $T_c$ nearly completely erases the pseudogap disorder before the turbulent disorder associated with the third (glassy) pjase transition sets in at T = $T_{OP}$.

Fig. 3. (a) This is Fig. 7(d) of [6], annotated to identify a region (circled ) of possible turbulence. This figure was constructed to show how two kinds of dopants could segregate to define metallic channels embedded in a pseudogapped matrix. (b) The simplest possible turbulent region (here a dark circle) arises at a junction where two metallic channels flow together.

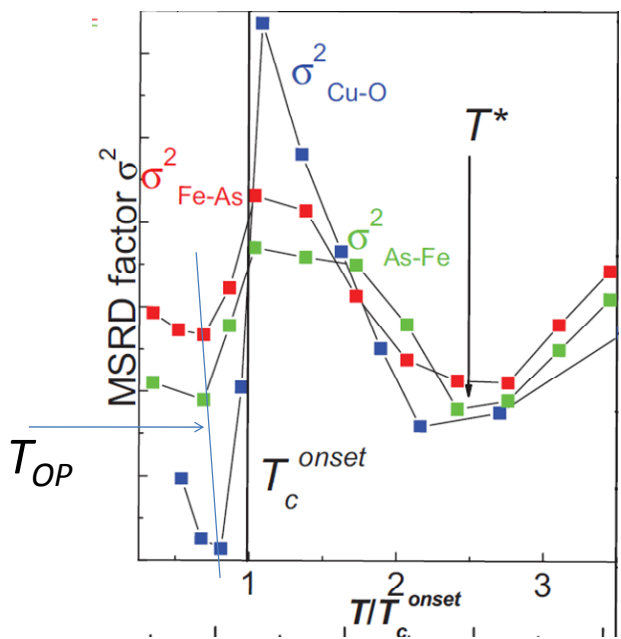

Fig.1

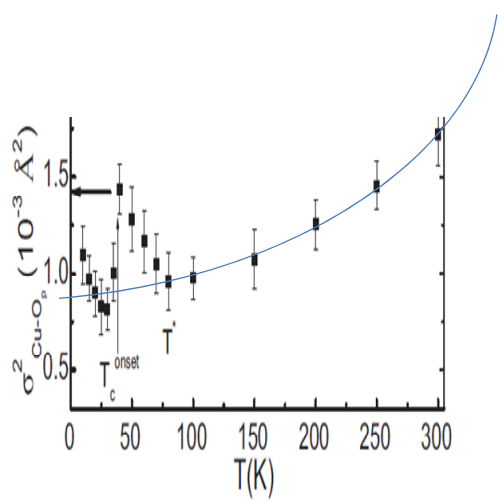

Fig. 2

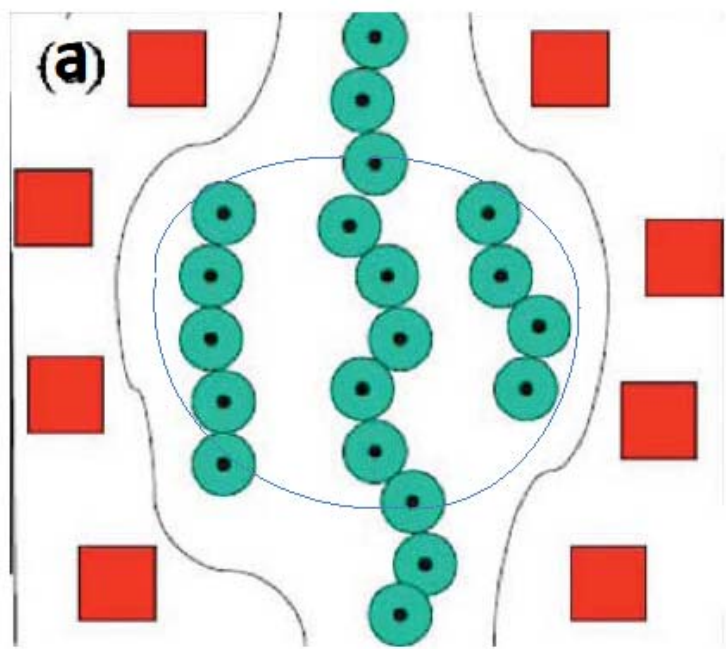

Fig. 3(a)

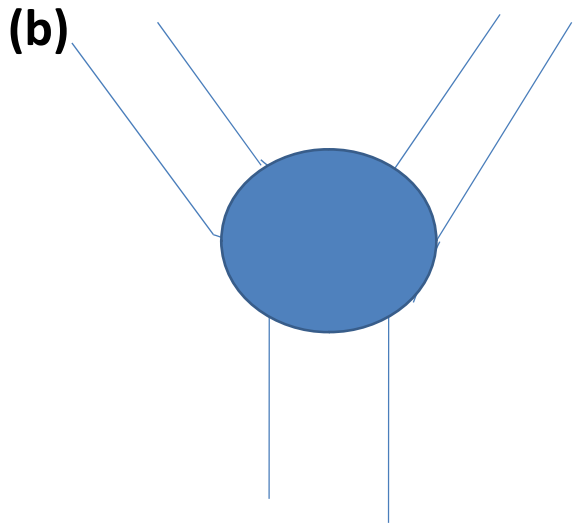

Fig. 3(b)